\documentclass[10pt,aps,prl,nofootinbib,twocolumn,noshowkeys,noshowpacs,longbibliography,
superscriptaddress,tightenlines,floatfix]{revtex4-2}
\usepackage{amsmath,amsfonts,amsthm,amssymb}
\usepackage{graphics,graphicx}
\usepackage{color}
\definecolor{darkblue}{RGB}{0,0,196}
\definecolor{darkgreen}{RGB}{0,120,0}

\def\beq{\begin{equation}}
\def\eeq{\end{equation}}
\def\st{\begin{equation}}
\def\stp{\end{equation}}
\def\ba{\begin{eqnarray}}
\def\ea{\end{eqnarray}}

\usepackage[colorlinks=true,linktocpage=true,linkcolor=darkblue,citecolor=red,urlcolor=darkblue]{hyperref}

\begin{document}
\preprint{}
 
    \title{Relativistic fluid dynamics in a `hydro' frame}
    
    %
\author{Sayantani Bhattacharyya}
    \email{sbhatta5@ed.ac.uk}
    \affiliation{School of Mathematics, University of Edinburgh, Peter Guthrie Tait Road, Edinburgh EH9 3FD, United Kingdom}
    \affiliation{School of Physical Sciences, National Institute of Science Education and Research, An OCC of Homi Bhabha National Institute, Jatni-752050, India}
    \author{Sukanya Mitra}
    \email{sukanya.mitra@niser.ac.in}
    \affiliation{School of Physical Sciences, National Institute of Science Education and Research, An OCC of Homi Bhabha National Institute, Jatni-752050, India}
    \author{Shuvayu Roy}
    \email{shuvayu.roy@iitgn.ac.in}
    \affiliation{School of Physical Sciences, National Institute of Science Education and Research, An OCC of Homi Bhabha National Institute, Jatni-752050, India}
    \affiliation{Indian Institute of Technology, Gandhinagar, Gujarat 382355, India}
\author{Rajeev Singh}
    \email{rajeevofficial24@gmail.com}
    \affiliation{School of Physical Sciences, National Institute of Science Education and Research, An OCC of Homi Bhabha National Institute, Jatni-752050, India}
    \affiliation{Department of Physics, West University of Timisoara, Bd.~Vasile P\^arvan 4, Timisoara 300223, Romania}
    \affiliation{Center for Nuclear Theory, Department of Physics and Astronomy, Stony Brook University, Stony Brook, New York, 11794-3800, USA}
	\date{\today} 
	\bigskip
\begin{abstract}
In this letter, we investigate how field redefinition influences the spectrum of linearized perturbations in relativistic fluid dynamics. We show that the hydrodynamic modes do not get affected under local field redefinition, whereas the non-hydrodynamic modes do. These non-hydrodynamic modes can be removed through a suitable all-order field redefinition. This process leads to a new frame containing only hydrodynamic modes, which we refer to as the `hydro' frame. Additionally, we demonstrate that the resulting stress-energy tensor may constitute an infinite series in momentum space, with the radius of convergence associated with the removed non-hydrodynamic mode, highlighting its role in the hydrodynamic expansion's validity. 
\end{abstract}
\date{\today}
	
\maketitle

%
\textit{Introduction:--}
In relativistic fluid dynamics, any viable model must satisfy critical physical criteria like causality and stability. These are typically assessed by examining linearized perturbations around an equilibrium background. When this background shows translational symmetry, the analysis uses Fourier modes, scrutinizing their spectrum to ensure compliance with physical constraints.
To ensure the robustness and observer-independence of these criteria, the analysis must respect the system's inherent symmetries. For instance, if the background fluid is rotationally invariant, all spatial coordinates related by global rotations must be treated equivalently. This means the observer can choose any coordinate set, and the derived constraints should remain consistent.
Incorporating rotational invariance into the perturbation analysis ensures this symmetry is reflected in the Fourier mode spectrum and resulting constraints. Thus, the fluid model's physical validity remains unaffected by the observer's coordinate choice, preserving its robustness and universality\cite{Hiscock:1985zz,Geroch:1990bw,Geroch:1995bx,Gavassino:2021kjm,Arnold:2014jva,Lehner:2017yes}.
Nevertheless, manifest invariance is not assured for all gauge freedoms in a model. When invariance is not explicit, it is important to separate artefacts of choice from the genuine physical spectrum to test stability and causality constraints effectively.

In relativistic hydrodynamics, defining fluid velocity and temperature beyond the perfect fluid case introduces ambiguity. In thermodynamic equilibrium, these variables are naturally defined by uniform thermodynamic potentials. Outside equilibrium, they can be redefined with arbitrary corrections that become nonzero with spacetime variations. 
These ambiguities are typically resolved by imposing constraints on the conserved currents, 
akin to gauge fixing in gauge theory, with common choices being the `Landau' or `Eckart' frame\cite{Israel:1976tn,Israel:1979wp}. 
However, one can always opt for constraints beyond these or even proceed without addressing these field ambiguities, as in BDNK theory\cite{Bemfica:2017wps,Kovtun:2019hdm,Bemfica:2019cop}. It appears that the spectrum of linearized perturbations is significantly altered when the fluid fields are redefined. The BDNK formalism, in particular, leaves the fluid frame unfixed to establish the frame in which the first-order fluid dynamics preserves causality and stability. This method diverges from the established difficulties related to causality and stability in the traditional Landau or Eckart frame\cite{Denicol:2008ha,Floerchinger:2017cii,Biswas:2020rps,Hoult:2023clg,Sammet:2023bfo,Domingues:2024pom}.
While rotational and translational symmetries are easily addressed in the analysis, gauge freedoms in defining fluid variables add further complexities.
Addressing these complexities is crucial for ensuring that the physical constraints derived from linearized perturbation analysis appropriately represent the underlying physics without any artefacts from arbitrary definitions\cite{Noronha:2021syv,Mitra:2023ipl,Bhattacharyya:2023srn,Bhattacharyya:2024tfj,Mitra:2024yei,Xie:2023gbo}.
It is well understood that the linearized spectrum changes with field redefinition, as the perturbations of the redefined fields correspond to different physical quantities than the original fields\cite{Grozdanov:2018fic,Grozdanov:2019uhi,Bea:2023rru,Hoult:2024cyx}. However, the causal nature of a specific fluid model should not be contingent upon the choice of variables used for its description. 

The causality of any relativistic fluid dynamic model depends on the nature of non-hydrodynamic modes. If the non-hydrodynamics modes are acausal, then the fluid model is also acausal. Thus, to establish a causal initial value problem for relativistic hydrodynamics, one must incorporate non-hydrodynamic modes\cite{Heller:2022ejw}.
These non-hydrodynamic modes are linked with the hyperbolicity of the equations, and field redefinitions may change the structure of non-hydrodynamic modes. It is possible that a fluid model that is causal in one frame might be acausal after the redefinition of fluid variables. Therefore, one may wonder which of the fluid models should be considered physical. 

We, in this work, provided a novel technique to find out which fluid model is physical under arbitrary field redefinition. The method is simple: we, through an all-order field redefinition, eliminate all non-hydrodynamic modes from the fluid model
and analyze the structure of the resulting modified energy-momentum tensor. This technique works provided the original stress-energy tensor (the one with non-hydrodynamic modes, before the field redefinition) has a finite number of terms in derivative expansion. We observe that if the new modified stress-energy tensor, after eliminating the non-hydrodynamic modes, incorporates higher derivative terms of the fluid variables that include up to infinite orders, then the fluid theory we started with is physical, whereas, if the modified stress-energy tensor turns out to have a finite number of terms, then the non-hydrodynamic modes of the original theory must emerge from field redefinition artefacts solely and should not be considered physical.
Another criterion we find is whether the dispersion polynomial neatly factorizes into hydrodynamic and non-hydrodynamic modes. If the non-hydrodynamic modes do not appear as distinct finite factors, then they are not frame artefacts\cite{Bhattacharyya:2024ohn}.
In this and a companion study\cite{Bhattacharyya:2024ohn}, we investigate the impact of field redefinition on the spectrum of linearized perturbations within relativistic fluid dynamics.
We have found that the spectrum of hydrodynamic modes (those with frequencies that vanish as spatial momentum approaches zero) remains unchanged by local field redefinition. In contrast, the spectrum of non-hydrodynamic modes (those with frequencies that approach a finite nonzero value as momenta diminish) is influenced by field redefinition.
We also show that non-hydrodynamic modes can be removed from the spectrum of linearized perturbations through a suitable, generally all-order field redefinition. Once this is applied, the spectrum will be simplified to include only the hydrodynamic modes of the original fluid model, and the stress-energy tensor becomes ready for analyzing the physical validity of the original non-hydrodynamic modes. We will refer to this new fluid frame as the ``hydro frame". Even if we start with a stress tensor that contains a finite number of terms, transforming to the ``hydro frame'' typically yields a stress tensor with an infinite number of terms, incorporating arbitrarily higher-order derivatives of the fluid variables.
In Ref.~\cite{Bhattacharyya:2024ohn}, we show how the non-hydrodynamic modes of the original theory could control the hydrodynamic expansion's validity of the new stress-energy tensor.

%
%
\smallskip
\textit{Setup:--} 
Let’s consider a set of variables $\{\Phi_i\}$ (such as velocity, temperature, and conserved charges in a fluid) governed by a system of nonlinear coupled PDEs represented as ${\cal E}(\{\Phi_i\})=0$ (for fluids, ${\cal E}$ includes equations from stress tensor and charge current conservation). We assume that $\{\Phi_i\} = \{\bar \Phi_i\}$ is a precise solution to ${\cal E}$ that is invariant under spacetime translations and spatial rotations. To study the spectrum, we first linearize the equations in ${\cal E}$ around $\{\bar\Phi_i\}$
\begin{eqnarray}
& \Phi_i^s=\bar \Phi_i + \epsilon\, \delta\Phi_i(\omega,k) e^{-i\omega t + i\vec k\cdot\vec x},\,\text{with}\,\, k = \sqrt{\vec k\cdot \vec k},\,\epsilon \ll 1\,,\nonumber\\
&{\cal E}(\{\Phi_i= \Phi_i^s\})=0 \Rightarrow \sum_j M_{ij}(\bar \Phi,\omega,k) \delta\Phi_j=0\,,
\label{lin1}
\end{eqnarray}
where $M_{ij}$ is the linearization matrix, and
the spectrum is obtained from the zeros of the determinant of $M_{ij}$. If the equation ${\cal E}$ contains a finite number of derivatives, then $\text{Det}[M]$ will be a finite polynomial in $\omega$ and $k$, resulting in a finite number of zeros.
Now consider a field redefinition $\Phi_i \rightarrow \Psi_i = \Phi_i + \Delta \Phi_i$, where the shift $\Delta \Phi_i$ is a nonlinear function of $\Phi_i$ and its derivatives. In the context of fluid dynamics, we assume that at equilibrium, $\Phi_i$ and $\Psi_i$ coincide, meaning $\Delta \Phi_i$ vanishes when evaluated at $\Phi_i = \bar \Phi_i$. Under this field redefinition, the equations of motion transform as ${\cal E}(\Phi) \rightarrow \tilde{\cal E}(\Psi)$. While the set of PDEs $\tilde{\cal E}$ can be fully determined from ${\cal E}$ and the transformation $\Delta \Phi$, it will have a distinctly different structure. We can linearize $\tilde{\cal E}$ to obtain the spectrum, similarly to how we did for ${\cal E}$ in~\eqref{lin1}, resulting in a different linearization matrix $\tilde M_{ij}$. Our objective is to establish the relationship between $M_{ij}$ and $\tilde M_{ij}$
\begin{eqnarray}
&\Psi_i^s = \bar \Phi_i + \epsilon\, \delta\Psi_i(\omega,k) e^{-i\omega t + i\vec k\cdot\vec x}\,\,\text{with} \,\, k = \sqrt{\vec k\cdot \vec k},\,\,\epsilon \ll 1\,,\nonumber\\
&\tilde{\cal E}(\{\Psi_i=\Psi_i^s\})=0 \Rightarrow \sum_j \tilde M_{ij}(\bar \Phi,\omega,k)\delta\Psi_j=0\,.
\label{lin2}
\end{eqnarray}
As $\Delta \Phi$ is vanishing when $\Phi = \bar \Phi$, it must be of order $\mathcal{O}(\epsilon)$ or higher for the field configuration in~\eqref{lin1}. Thus, $\Delta \Phi$ can be represented by the matrix equation up to ${\cal O}(\epsilon^2)$
\begin{equation}
\Delta\Phi_i (\Phi^s)= \epsilon\sum_j S_{ij}(\bar\Phi,\omega,k) \delta\Phi_j(\omega,k) e^{-i\omega t + i\vec k\cdot\vec x}\,.
\label{shiftlin}
\end{equation}
Under field redefinition, $\Phi_i^s\rightarrow \Psi_i^s$
\begin{eqnarray}
\Psi_i^s &=& \bar \Phi_i + \epsilon~ \delta\Psi_i(\omega,k) e^{-i\omega t + i\vec k\cdot\vec x}\,,\nonumber \\
&=& \Phi_i^s + \Delta\Phi_i (\Phi^s)\,, \label{translin2}\\
&=& \bar \Phi_i + \epsilon\sum_j\left[\delta_{ij} + S_{ij}(\bar\Phi,\omega,k)\right]\delta\Phi_j(\omega,k) e^{-i\omega t + i\vec k\cdot\vec x}\,,\nonumber\\
&&{\rm implying}\nonumber\\
&&\delta\Psi_i(\bar\Phi,\omega,k) = \sum_j\left[\delta_{ij} + S_{ij}(\bar\Phi,\omega,k)\right]\delta\Phi_j(\bar\Phi,\omega,k).\qquad
    \label{eq:lin1.2}
\end{eqnarray}
Putting \eqref{eq:lin1.2} in \eqref{lin2} and comparing with \eqref{lin1}, we obtain
\begin{eqnarray}
\sum_{jk} \tilde M_{ij} &&\left(\delta_{jk} + S_{jk}\right)\delta\Phi_k =0
\Rightarrow  M_{ik} = \sum_{j} \tilde M_{ij} \left(\delta_{jk} + S_{jk}\right)\nonumber\\
&&\Rightarrow {\rm Det}[M] = {\rm Det}[\tilde M] \, {\rm Det}[ {\bf 1} + S].
\label{finclude}
\end{eqnarray}
The zeros of ${\rm Det}[{\bf 1} + S]$ represent new modes in the $\Phi$ frame that are absent in the $\Psi$ frame. These new zero modes arise from perturbations in the $\Phi$ frame, where the additional terms coming from the linearized field redefinitions are exactly canceled by the linearized fluctuations. In the $\Psi$ frame, these new modes indicate no fluctuations, meaning that they are artefacts of the frame transformation with no physical significance. Importantly, depending on the choice of field redefinitions, these artificial modes may appear unstable or acausal, even if the theory in the $\Psi$ frame is entirely valid. However, in the $\Phi$ frame, there is no straightforward way to distinguish these artefacts from genuine physical modes. 
%

\smallskip
\textit{Within the framework of relativistic fluid dynamics:--}
Here, we concentrate on fluid variables to establish the most general form of the frame transformation matrix $S$ (see eq.~\eqref{shiftlin}). To simplify our analysis, we will focus on uncharged fluids, where fluid velocity and temperature are the only variables, and the equation of motion is dictated by stress tensor conservation.

Let the velocity and temperature in two different frames, `Frame-1' and `Frame-2,' be denoted as $\{\hat u^\mu, \hat T\}$ and $\{u^\mu, T\}$, respectively. The shift functions $\Delta u^\mu$ and $\Delta T$ are defined as
\begin{equation}\label{fluidframe}
u^\mu = \hat u^\mu +\Delta u^\mu (\hat u,\hat T)\,, \qquad T = \hat T + \Delta T(\hat u, \hat T)\,.
\end{equation}
Fluid variables in different frames are expected to align in a rotationally and translationally invariant equilibrium. Consequently, every term in $\Delta u^\mu$ and $\Delta T$ must include at least one spacetime derivative, ensuring that the shift variables are non-zero only for spatially or temporally non-uniform fluid profiles. We focus on terms that contribute at the linear order in fluctuation amplitude when evaluated on equilibrium fluid profiles plus small fluctuations. This holds for single derivative terms, while for multiple derivatives, all must act on a single fluid variable such as $(u^\alpha u^\beta\partial_\alpha\partial_\beta)u^\mu$.
Therefore, the most general expressions for $\Delta u^\mu$ and $\Delta T$ that may influence the spectrum of small fluctuations are\cite{Bhattacharyya:2024ohn}
\begin{eqnarray}
&\Delta u^\mu=F_u (\hat u\cdot\partial) \hat u^\mu + F_T (\frac{\hat P^{\mu\alpha}\partial_\alpha \hat T}{\hat T})+ R_u (\hat P^{\mu\theta}\hat P^{\alpha\beta} \partial_\alpha\partial_\beta \hat u_\theta)\,,\nonumber\\
&\frac{\Delta T}{\hat  T}=G_u (\partial\cdot \hat u) + G_T (\frac{\hat  u^\alpha\partial_\alpha\hat T}{\hat T})+ R_T (\frac{\hat P^{\alpha\beta} \partial_\alpha\partial_\beta \hat T}{\hat T}),
\label{motstg}
\end{eqnarray}
where $F_{u(T)} = F_{u(T)}[(\hat u\cdot\partial),\hat P^{\alpha\beta}\partial_\alpha\partial_\beta]$, $G_{u(T)} = G_{u(T)}[(\hat u\cdot\partial),\hat P^{\alpha\beta}\partial_\alpha\partial_\beta]$ are
linear differential operators
\begin{eqnarray}\label{definc}
F_{u(T)} &\equiv \sum_{m,n} f_{m,n}^{u(T)} \left[\hat u\cdot\partial\right]^m \left[\hat P^{\alpha\beta}\partial_\alpha\partial_\beta\right]^n\,, \\
G_{u(T)} &\equiv \sum_{m,n} g^{u(T)}_{m,n} \left[\hat u\cdot\partial\right]^m \left[\hat P^{\alpha\beta}\partial_\alpha\partial_\beta\right]^n\,,\nonumber
\end{eqnarray}
and $R_{u(T)}=R_{u(T)}[\hat P^{\alpha\beta}\partial_\alpha\partial_\beta]$ do not depend on the operator $(\hat u\cdot\partial)$.
They are included to account for field redefinitions that remain non-zero as $\omega \rightarrow 0$. These can also be expanded in the small $k$ limit as
$R_{u(T)} =  \sum_{m} r^{u(T)}_m  \left[\hat P^{\alpha\beta}\partial_\alpha\partial_\beta\right]^m$.
Here, $f_{m,n}^{u(T)}$, $g^{u(T)}_{m,n}$, and $r^{u(T)}_m$ are functions of temperature, and $\hat{P}^{\mu\nu}$ is the projector orthogonal to $\hat{u}^\mu$, $\hat{P}^{\mu\nu} = \eta^{\mu\nu} + \hat{u}^\mu \hat{u}^\nu$. In the context of linearized analysis, the derivatives $(\hat{u}^\alpha \partial_\alpha)$ and $(\hat{P}^{\alpha\beta} \partial_\alpha)$ commute, so their ordering in the functions don't matter.
For a rotational and translationally invariant equilibrium case, we have
 $\bar u^\mu = \{1,0,0,0\}, \quad \bar T = {\rm constant}$,
where the perturbed solution is
$u_\mu^{(s)} = \bar u_\mu +\epsilon~ \delta u_\mu \,e^{-i\omega t + i\vec k\cdot\vec x},\quad T^{(s)} = \bar T +\epsilon\, \delta T \,e^{-i\omega t + i\vec k\cdot\vec x}$.

As $u_\mu^{(s)}$ and $\bar u^\mu$ are normalized to unity, hence, $\delta u\cdot \bar u \sim {\cal O}(\epsilon^2)$ or $\delta u^\mu$ must be of the form
 $\delta u^\mu =\{0,\vec\beta\}$ where $\vec \beta$ reads
$\vec\beta = \beta_k\left(\vec k/ k\right)+\vec \beta_\perp$ with $k = \sqrt{\vec k\cdot\vec k}$,  $\beta_k =\left(\vec\beta\cdot\vec k\right)/ k$, and $\vec k\cdot\vec\beta_\perp =0$.
 
After putting in Eq.~\eqref{motstg} we obtain
$\Delta u_\mu = \{0, \vec {\Delta u}\}$ with $\vec {\Delta u}=\Delta u_k \left(\vec k/ k\right) +\vec{ \Delta u_\perp}$
such that $\vec{\Delta u}_\perp\cdot\vec k =0$.
Thus, the shift functions we receive are\cite{Bhattacharyya:2024ohn}
\begin{gather}\label{flufra}
 \begin{bmatrix} \Delta T\\
 \Delta u_k\\
 \Delta u_\perp
 \end{bmatrix}
 = S_{ij} \,
    \begin{bmatrix}
    \delta T\\
    \beta_k\\
    \beta_\perp
    \end{bmatrix}\,,
   \end{gather}
   with the matrix (as mentioned in~\eqref{shiftlin})
       \begin{footnotesize}
\begin{equation}
S_{ij}
 =\begin{bmatrix}
  -i\omega\, G_T-k^2R_T&ik \,G_u&0\\
  ik\, F_T &-i\omega F_u-k^2 R_u&0\\
  0&0&-i\omega F_u-k^2R_u
   \end{bmatrix}\nonumber
   \end{equation}
       \end{footnotesize}
where $\Delta u_k$ and $\Delta u_\perp$ are defined the same way as $\beta_k$ and $\beta_\perp$. Now we compute ${\rm Det}\left[{\mathbf{1}} + S\right]$ as
   \begin{eqnarray}
   &{\cal F}(\omega, k^2)\equiv {\rm Det}[{\mathbf{1}} +S] =(1-i\omega F_u-k^2R_u)\times \nonumber\\
   &\left[(1-i\omega G_T-k^2R_T)(1-i\omega F_u-k^2R_u) + k^2 F_T\,G_u\right].\quad
   \label{fluidet}
   \end{eqnarray}
We have suppressed the arguments of $F_{u(T)}$, $G_{u(T)}$, and $R_{u(T)}$ for simplicity. From Eq.~\eqref{finclude}, we note that a general field redefinition of the fluid variables introduces an additional factor, $\mathcal{F}$, to the dispersion polynomial which is a polynomial in $\omega$ and $k^2$. This generates new modes in the system.
Some remarks on the properties of $F_{u(T)}$ and $G_{u(T)}$: If $F_u$ and $G_T$ are vanishing, and $F_T$ and $G_u$ do not depend on the operator $(u\cdot\partial)$ or $(-i\omega)$ in Fourier space, then $\mathcal{F}$ will have no zeros for any $\omega$, resulting in no new modes. However, depending on $F_T$ and $G_u$, $\mathcal{F}$ might have zeros for some real values of $k$. If every term in $F_u$ and $G_T$ includes at least one factor of $(P^{\mu\nu}\partial_\mu\partial_\nu)$ or $k^2$ in Fourier space, the frequency of these new modes will diverge as $k \rightarrow 0$. If either $F_u$ or $G_T$ has at least one term without $(P^{\mu\nu}\partial_\mu\partial_\nu)$, at least one new mode will appear with a finite, non-zero frequency as $k \rightarrow 0$, resembling a genuine ``non-hydrodynamic'' mode. None of the zeros of $\mathcal{F}$ will have the form $\lim_{k \rightarrow 0} \omega(k) = 0$. Thus, frame redefinitions will not
generate new hydrodynamic modes, and redefining fluid variables will not affect the spectrum of hydrodynamic modes in Fourier space.

   %
   \smallskip
\textit{Removal of mode from the spectrum via field redefinition:--}
In the previous section, we noticed that `Frame-2' dispersion polynomial has more non-hydrodynamic modes (coming from the zeros of $\mathcal{F}$) than in `Frame-1' due to frame transformation.

Now, let's reverse this process—starting with the `Frame-2' equations (using the hatted fluid variables $\hat{u}^\mu$ and $\hat{T}$) and implementing the inverse transformations from~\eqref{fluidframe} and \eqref{motstg} to return to `Frame-1' equation with fluid variables $u^\mu$ and $T$. 
This inversion would naturally eliminate the extra factor $\mathcal{F}$ from the `Frame-2' dispersion polynomial. From the viewpoint of `Frame-2', this effectively removes a non-hydrodynamic mode from the spectrum through field redefinition.
The possibility to fully absorb a non-hydrodynamic mode through a field redefinition implies that such a mode may not be physical. A truly physical non-hydrodynamic mode should never be completely erased by a frame transformation. 

In this section, we will investigate an ``inverse transformation'' to remove a mode from the spectrum and analyze if physical non-hydrodynamic modes persist as expected.
The dispersion polynomials in `Frame-1' and `Frame-2', indicated as $P(\omega,k^2)$ and $\hat P(\omega, k^2)$, respectively, are related as
\begin{eqnarray}
    \hat P(\omega, k^2) = P(\omega,k^2){{\cal F}(\omega,k^2)},
\end{eqnarray}
where $\mathcal{F}$ is given in \eqref{fluidet}, with
\begin{eqnarray}
F_{u(T)} &\equiv& \sum_{m,n} f_{m,n}^{u(T)} \left[-i\omega\right]^m \left[-k^2\right]^n,\nonumber\\
G_{u(T)} &\equiv& \sum_{m,n} g^{u(T)}_{m,n} \left[-i\omega\right]^m \left[-k^2\right]^n,\nonumber\\
R_{u(T)} &\equiv& \sum_{m} r^{u(T)}_{m} \left[-k^2\right]^m\,.
\label{rel12}
\end{eqnarray}
Due to the isotropy of the background fluid profile and the constitutive relations, the linearized dynamics in the shear and sound sectors will get disentangled. Consequently, both $\hat{P}(\omega, k^2)$, $P(\omega, k^2)$ and $\mathcal{F}(\omega, k^2)$ must factor into contributions from the shear and sound sectors
\begin{eqnarray}
\hat P_{\rm sh}(\omega, k^2) &=& P_{\rm sh}(\omega, k^2) (1-i\omega F_u-k^2R_u)\,,\label{shearsound}\\
\hat P_{{\rm snd}}(\omega,k^2) &=& P_{{\rm snd}}(\omega, k^2)\big[(1-i\omega G_T-k^2R_T)\nonumber\\
&&(1-i\omega F_u-k^2R_u) + k^2 F_T \,G_u\big],\nonumber
\end{eqnarray}
where $\hat P(\omega, k^2) = \hat P_{\rm sh}(\omega, k^2)\hat P_{{\rm snd}}(\omega,k^2)$, and
$P(\omega,k^2) = P_{\rm sh}(\omega, k^2) P_{{\rm snd}}(\omega, k^2)$.
Assume that $P_{\rm sh}(\omega, k^2)$ has $N^{\rm sh}_1$ hydrodynamic modes with frequencies $\omega_{\rm sh}^a(k)$ and $N^{\rm sh}_2$ non-hydrodynamic modes with $\mathfrak{w}_{\rm sh}^a(k^2)$ where, respectively, $a = \{1, 2, \ldots, N^{\rm sh}_1\}$  and $a = \{1, 2, \ldots, N^{\rm sh}_2\}$. 
And in the sound channel, it has $N^{\rm snd}_1$ hydrodynamic modes with $\omega_{\rm snd}^a(k)$  $\left(a = \{1, 2, \ldots, N^{\rm snd}_1\}\right)$ and $N^{\rm snd}_2$ non-hydrodynamic modes with $\mathfrak{w}_{\rm snd}^a(k^2)$ $\left(a = \{1, 2, \ldots, N^{\rm snd}_2\} \right)$. Thus, $P_{\rm sh}(\omega,k^2)$ and $P_{\rm snd}(\omega, k^2)$ can be factored as
$P_{\rm sh}(\omega,k^2) =\left[\prod_{a=1}^{N^{\rm sh}_1}(\omega-\omega_{\rm sh}^{(a)})\right]\left[\prod_{a=1}^{N^{\rm sh}_2}(\omega-{\mathfrak w}_{\rm sh}^{(a)})\right]$, and
$P_{{\rm snd}}(\omega,k^2) =\left[\prod_{a=1}^{N^{{\rm snd}}_1}(\omega-\omega_{{\rm snd}}^{(a)})\right]\left[\prod_{a=1}^{N^{{\rm {\rm snd}}}_2}(\omega-{\mathfrak w}_{{\rm {\rm snd}}}^{(a)})\right]$.
Both $\omega^{(a)}_{{\rm sh}/{\rm snd}}$ and $\mathfrak{w}^{(a)}_{{\rm sh}/{\rm snd}}$ are generally complex, non-polynomial functions of $k$ with limits:
$\lim_{k\rightarrow0}\omega^{(a)}_{{\rm snd}/{\rm sh}}(k)=0, \quad \lim_{k\rightarrow0}{\mathfrak w}^{(a)}_{{\rm sh}/{\rm snd}}(k)=c^{(a)}_{{\rm sh}/{\rm snd}} \, \forall a$.

Our aim is to identify a frame transformation that eliminates non-hydrodynamic modes while remaining precisely equivalent to the original system in the hydrodynamic sector. This exact equivalence requires that both the frequencies and eigenvectors in the hydrodynamic sector align perfectly before and after the transformation.
It's essential to recognize that the permissible non-trivial profiles of linearized perturbations also cleanly factorize into two subspaces: the 2D space of the sound channel, characterized by the velocity perturbation aligned with the wave vector ($\vec{k}$) and the temperature perturbation, and the 1D subspace of the shear channel, represented by the velocity perturbation perpendicular to $\vec{k}$.
As the hydrodynamic eigenspace in the shear channel is one-dimensional, the frame transformation will automatically match the eigenvector after removing the shear non-hydrodynamic frequency from the spectrum without impacting the hydrodynamic frequency. In contrast, in the sound channel, the eigenvector is a specific vector in the two-dimensional subspace of temperature and longitudinal velocity perturbations, defined by the ratio $\mathcal{R}$ of these two components: ${\cal R}(\omega,k) = {\delta T}/{\delta u_k}$.
In the sound channel, after eliminating the non-hydrodynamic mode, we must ensure that both the hydrodynamic frequencies and $\mathcal{R}$ remain unchanged before and after the frame transformation.
Hence, we obtain
\begin{eqnarray}\label{eq:frametrans}
&{(1-i\omega F_u-k^2R_u)}\left[\prod_{a=1}^{N^{\rm sh}_2}(\omega-{\mathfrak w}_{\rm sh}^a)\right]=C_{\rm sh}\,,\\
&\left[(1-i\omega G_T-k^2R_T)(1-i\omega F_u-k^2R_u) + k^2 F_T\,G_u\right]\nonumber\\
&\times \left[\prod_{a=1}^{N^{{\rm {\rm snd}}}_2}(\omega-{\mathfrak w}_{{\rm {\rm snd}}}^a)\right]=C_{{\rm {\rm snd}}},\nonumber
\end{eqnarray}
where $C_{\rm sh} = \lim_{k\rightarrow 0} \prod_{a=1}^{N_2^{\rm sh}}\left[-{\mathfrak w}^a_{\rm sh}\right]$ and $C_{{\rm {\rm snd}}} = \lim_{k\rightarrow 0}\prod_{a=1}^{N_2^{{\rm {\rm snd}}}}\left[-{\mathfrak w}^a_{{\rm {\rm snd}}}\right]$.
If $F_{u(T)}$, $G_{u(T)}$, and $R_{u(T)}$ are finite polynomials in $\omega$ and $k^2$, the condition, \eqref{eq:frametrans}, cannot be fulfilled. However, by permitting an infinite-order derivative expansion in Fourier space (around $\omega=k=0$), we can, in principle, eliminate all non-hydrodynamic modes from the dispersion polynomial through an appropriate selection of expansion coefficients $f_{m,n}^{u(T)}$, $g^{u(T)}_{m,n}$, $R^{u(T)}_m$. Moreover, the radius of convergence for this infinite-order frame transformation in Fourier space will be dictated by the position of the nearest non-hydrodynamic mode to the origin in the multi-dimensional space of real momenta and complex frequencies.
Let’s normalize the perturbed solution in the sound channel as:
$u_\mu^{(s)} = \bar u_\mu +\epsilon\, \{0,{\vec k/ k}\} \,e^{-i\omega t + i\vec k\cdot\vec x}$, $T^{(s)} = \bar T +\epsilon\, \delta T \,e^{-i\omega t + i\vec k\cdot\vec x}$, where $\delta T$ is dictated by ${\cal R}(\omega,k)$ before the frame transformation and $\hat {\cal R}(\omega,k)$ after the transformation. The equivalence of the hydrodynamic sector requires that:
\begin{equation}\label{eq:frameeigen}
{\cal R}(\omega^a_{{\rm {\rm snd}}}, k) = \hat{\cal R}(\omega^a_{{\rm {\rm snd}}},k) \quad \forall a = \{1,2,\dots , N_1^{{\rm {\rm snd}}}\}\,.
\end{equation}
Let’s examine whether we can obtain a solution for this set of coupled nonlinear equations (\eqref{eq:frametrans} and \eqref{eq:frameeigen}) and to what degree the solution would be unique.
Naively, it appears that there are $(N_1^{{\rm snd}} + 2)$ equations for six unknown functions. However, $R_T$ and $R_u$ depend only on $k$. Taking the limit $\omega \rightarrow 0$ in the two equations of \eqref{eq:frametrans} effectively yields two additional equations involving $R_T$, $R_u$, and $[F_T\,G_u]_{(0)}$ (defined as $[F_T\,G_u]_{(0)} \equiv \lim_{\omega \rightarrow 0} [F_T(\omega,k) G_u(\omega,k)]$). 

Using these additional equations, we can uniquely determine $R_T$ and $R_u$ in terms of $[F_T ~ G_u]_{(0)}$ and other known functions of $k$, such as ${\mathfrak w}^a_{{\rm snd}}(k)$ and ${\mathfrak w}^a_{\rm sh}(k)$. Next, we substitute the solutions for $R_T$ and $R_u$ and reintroduce $\omega$ into \eqref{eq:frametrans}. We can now uniquely solve for $F_u$ and $G_T$ in terms of the product $F_T G_u$ and $[F_T G_u]_{(0)}$.
At this point, both equations in \eqref{eq:frametrans} will be solved for all values of $\omega$ and $k$. Next, we have $N_1^{{\rm snd}}$ equations from \eqref{eq:frameeigen} to determine $F_T$ and $G_u$. For an uncharged hydrodynamic theory like those discussed here, $N_1^{{\rm snd}}=2$, giving us two equations for the two remaining unknowns, $F_T$ and $G_u$. However, the equations in \eqref{eq:frameeigen} are evaluated at $\omega = \omega^a_{{\rm snd}}(k)$, which prevents us from uniquely determining $F_T$ and $G_u$ for all values of $\omega$. In particular, we can add two arbitrary functions of the product $\prod_a(\omega - \omega^a_{{\rm snd}})$ to any solutions for $F_T$ and $G_u$, and they will still satisfy \eqref{eq:frameeigen}. 
Despite the minor ambiguity in $F_T$ and $G_u$ noted above, we can nearly uniquely determine the functions in the linearized frame transformation by imposing that the spectrum in the transformed frame contains only the hydrodynamic modes\cite{Bhattacharyya:2024ohn}.

\smallskip
\textit{Field redefinition and non-hydrodynamic mode in BDNK:--} 
Let us now apply the above algorithm to eliminate the non-hydrodynamic modes in the BDNK theory\cite{Bemfica:2017wps,Kovtun:2019hdm}. 
The stress-energy tensor for an uncharged conformal system reads
\begin{eqnarray}
T^{\mu\nu} &= T^4\big\{(4 u^\mu u^\nu + \eta^{\mu\nu})
 +   4  \big[ \theta \left(u^\mu q^\nu + u^\nu q^\mu\right) \nonumber\\
 &+ \chi \,E (4 u^\mu u^\nu + \eta^{\mu\nu})- 2 \lambda \, \sigma^{\mu\nu} \big]\big\}\,,
 \label{bdnkstress}
\end{eqnarray}
with
$q^\mu\equiv (u\cdot\partial)u^\mu +  \left(P^{\mu\nu} \partial_\nu T\over T\right)$,
$E \equiv {(u\cdot \partial) T\over T} + {\left(\partial\cdot u\right)}/{3}$, and
$\sigma^{\mu\nu} \equiv \left({P^{\mu\alpha} P^{\nu\beta} +P^{\mu\beta} P^{\nu\alpha} \over 2 }-{P^{\mu\nu}P^{\alpha\beta}\over 3} \right)\left(\partial_\alpha u_\beta\right)$. Here, $\{\chi(T),\theta(T),\lambda(T)\}$ are constants.
The dispersion polynomial is\footnote{We have re-scaled the transport coefficients ($\theta$, $\chi$, and $\lambda$) by proper factors of $T$ so that the products of these coefficients with frequency ($\omega$) or momentum ($k$) are dimensionless.}
\begin{eqnarray}
&&P_\text{BDNK}(\omega,k) =
 \left[\omega  (1- i\chi\, \omega )+i\lambda k^2  \right] \times \nonumber\\
 &&\big[\left[3 \omega 
   (1-i\theta \omega)-i\chi k^2\right] \left[3\omega  (1-i\chi \omega )-i(\theta-4\lambda) k^2\right]\nonumber\\
   &&\quad +3i k^2 [1-i(\chi+\theta) \omega ]^2 \big].\label{eq:disperseBDNK}
\end{eqnarray}
It has one hydrodynamic and one non-hydrodynamic mode in the shear channel, and two hydrodynamic and two non-hydrodynamic modes in the sound channel. To determine the frame transformation as an expansion in $\omega$ and $k$, it suffices to know the modes in a similar expansion in $k$ (up to ${\cal O}(k^4)$)
\begin{eqnarray}
&{\mathfrak w}_{\rm sh}=-{i\over\theta} + i\lambda~ k^2 \,, \quad \omega_{\rm sh}=-i\lambda~k^2 \,,\nonumber\\
&{\mathfrak w}^{(1)}_{{\rm snd}}=-{i\over\theta} + {i\over 3}\left[4\lambda -{\theta\chi\over(\theta-\chi)}\right]k^2 \,,\nonumber\\
&{\mathfrak w}^{(2)}_{{\rm snd}}=-{i\over\chi} + {i\over 3}\left(\theta\chi\over\theta-\chi\right)k^2 \,,\label{modesBDNK}\\
&\omega^{(1)}_{{\rm snd}}={k\over\sqrt{3}}- \left(2i\lambda\over 3\right) k^2 - \left(2\lambda^2\over 3\sqrt{3}\right) k^3 \,,\nonumber\\
&\omega^{(1)}_{{\rm snd}}=-{k\over\sqrt{3}}- \left(2i\lambda\over 3\right) k^2 + \left(2\lambda^2\over 3\sqrt{3}\right) k^3 \,.\nonumber
\end{eqnarray}
Then using equations \eqref{eq:frametrans} and \eqref{eq:frameeigen}, we solve for $F_{u(T)}$, $G_{u(T)}$, and $R_{u(T)}$.
Taking $\omega\rightarrow 0 $ limit  in the 1st equation of \eqref{eq:frametrans}, we obtain the solution for $R_u(k)$
\begin{equation}\label{sol:ru}
R_u(k)=\frac{1}{k^2} \left( {\mathfrak w}_{\rm sh} +{i\over\theta}\over {\mathfrak w}_{\rm sh}\right).
\end{equation}
Now we put \eqref{sol:ru} in \eqref{eq:frametrans} keeping $\omega$ finite to solve $F_u(\omega,k)$
\begin{equation}\label{sol:fu}
F_u(\omega,k)=-{1/\theta\over {\mathfrak w}_{\rm sh}(\omega-{\mathfrak w}_{\rm sh})}.
\end{equation}
Then we turn to the second equation of \eqref{eq:frametrans} and use Eqs.~\eqref{sol:ru} and \eqref{sol:fu}.
Again, taking $\omega\rightarrow 0$ limit of the resulting equation, we receive the solution for $R_T(k)$
\begin{equation}
R_T(k)=i\theta \,{\mathfrak w}_{\rm sh}\left[F_T\,G_u\right]_{(0)}+\frac{1}{k^2} \left[1+{i\over\chi} \left({\mathfrak w}_{\rm sh}\over {\mathfrak w}_{{\rm snd}}^{(1)}{\mathfrak w}_{{\rm snd}}^{(2)}\right)\right].
\label{sol:rt}
\end{equation}
Finally, putting \eqref{sol:rt} in the same equation with $\omega$ finite, we solve for $G_T(\omega,k)$
\begin{eqnarray}
&G_T(\omega,k)=\theta\,{\mathfrak w}_{\rm sh}\left(k^2\over\omega\right)\left(F_T G_u -\left[F_T G_u\right]_{(0)}\right)\label{sol:gt}\\
&-\theta \,k^2(F_T \,G_u)-\frac{1}{\chi} \left[{\mathfrak w}_{{\rm snd}}^{(1)}{\mathfrak w}_{{\rm snd}}^{(2)}+{\mathfrak w}_{\rm sh}\left(\omega -{\mathfrak w}_{{\rm snd}}^{(1)}-{\mathfrak w}_{{\rm snd}}^{(2)}\right)\over {\mathfrak w}_{{\rm snd}}^{(1)}{\mathfrak w}_{{\rm snd}}^{(2)}\left(\omega -{\mathfrak w}_{{\rm snd}}^{(1)}\right)\left(\omega -{\mathfrak w}_{{\rm snd}}^{(2)}\right)\right].\nonumber
\end{eqnarray}
We have two unknowns, $F_T(\omega, k)$ and $G_u(\omega, k)$, and two equations~\eqref{eq:frameeigen} evaluated at $\omega_{{\rm snd}}^{(1)}$ and $\omega_{{\rm snd}}^{(2)}$. First we determine $\hat{\cal R}(\omega,k)$ and ${\cal R}(\omega,k)$ as
\begin{eqnarray}
&\hat{\cal R}(\omega,k) ={ k\left[G_u\left(\theta k^2  +3\chi\omega^2 + 3 i\omega \right) +i\left(1 - i\omega F_u-k^2R_u\right)\left(i + \theta\omega+\chi\omega\right)\right]\over k^2\left(i + \theta\omega+\chi\omega\right)F_T + i\left(1-k^2R_T -i\omega G_T\right)\left(\theta k^2  +3\chi\omega^2 + 3 i\omega \right)},\nonumber\\
&{\cal R}(\omega,k) =  k\left[i + (\theta+\chi)\omega\over k^2\theta +3\omega (i +\chi\omega)\right],\nonumber\\
&{\cal R}(\omega_{{\rm snd}}^{(1)},k) =\hat{\cal R}(\omega_{{\rm snd}}^{(1)},k), \quad 
{\cal R}(\omega_{{\rm snd}}^{(2)},k) =\hat{\cal R}(\omega_{{\rm snd}}^{(2)},k).\,\,\,\,~~
\label{pres}
\end{eqnarray}
Note that \eqref{pres} must be fulfilled only at the frequencies of the hydrodynamic sound modes, $\omega^{(1)}_{{\rm snd}}(k)$ and $\omega^{(2)}_{{\rm snd}}(k)$. 
Solving these equations does not determine the unknown functions $F_T$ and $G_u$ for all $\omega$ and $k$. However, our aim is to demonstrate the existence of at least one frame transformation that eliminates the non-hydrodynamic modes. By assuming $F_T$ and $G_u$ do not depend on $\omega$, equation \eqref{pres} uniquely solves these functions, providing a suitable frame transformation. With this assumption, the solutions for $R_T$ and $G_T$ simplify
\begin{eqnarray}
&R_T(k)=i\theta \,{\mathfrak w}_{\rm sh}F_T G_u+\frac{1}{k^2} \left[1+{i\over\chi} \left({\mathfrak w}_{\rm sh}\over {\mathfrak w}_{{\rm snd}}^{(1)}{\mathfrak w}_{{\rm snd}}^{(2)}\right)\right],\label{rtgtsimp}\\
&G_T(\omega,k)=-\theta \,k^2(F_T G_u)
\nonumber\\
&\qquad \qquad \qquad -{1\over\chi} \left[{\mathfrak w}_{{\rm snd}}^{(1)}{\mathfrak w}_{{\rm snd}}^{(2)}+{\mathfrak w}_{\rm sh}\left(\omega -{\mathfrak w}_{{\rm snd}}^{(1)}-{\mathfrak w}_{{\rm snd}}^{(2)}\right)\over {\mathfrak w}_{{\rm snd}}^{(1)}{\mathfrak w}_{{\rm snd}}^{(2)}\left(\omega -{\mathfrak w}_{{\rm snd}}^{(1)}\right)\left(\omega -{\mathfrak w}_{{\rm snd}}^{(2)}\right)\right].\nonumber
\end{eqnarray}
Even after assuming $F_T$ and $G_u$ are independent of $\omega$, eq.~\eqref{pres} remains a set of coupled nonlinear algebraic equations, with nonlinearity due to $R_T$ depending on the product of $F_T$ and $G_u$. While solving is straightforward, the exact expressions for $F_T$ and $G_u$ are cumbersome compared to other functions $F_u$, $R_u$, $R_T$, and $G_T$. The complete details are presented in\cite{Bhattacharyya:2024ohn}. The final leading solutions, up to $\mathcal{O}(k^2)$, are as follows:
\begin{equation}
G_u(k)=g_0 \,,\quad
  F_T(k)=3g_0 + \chi-\theta.
  \label{ftgu}
\end{equation}
with $g_0 \equiv -(1/6) (\sqrt{5 \theta^2+16 \lambda (\lambda - \theta)+8 \lambda  \chi -3 \chi(\chi+2\theta)}-\theta -4
   \lambda +\chi )$.
After solving for all the functions in the frame transformation in Fourier space, we can convert them to position space using the following simple replacement:
$\omega\rightarrow-i (u\cdot\partial), \quad k^2\rightarrow -P^{\mu\nu}\partial_\mu\partial_\nu$.

At this point, we note that although we provided formal exact solutions for $F_u$, $R_u$, $R_T$, and $G_T$, these should be understood as expansions in non-negative powers of $\omega$ and $k$. Polynomials of $k$ and $\omega$ in the denominator imply an infinite power series with a convergence radius computed by the zeros of these polynomials.
Our study relies on the derivative expansion, and replacing $\{\omega, k^2\}$ with derivatives doesn't make sense if $\omega$ and $k$ appear in the denominator. All functions $F_{u(T)}$, $G_{u(T)}$, and $R_{u(T)}$ remain finite as $\omega \to 0$ and/or $k \to 0$. This ensures no negative powers of $(u \cdot \partial)$ or $(P_\mu^\nu \partial_\nu)$ arise after expansion and replacement.
Furthermore, the final frame transformation solution must depend only on even powers of $k$ to apply the replacement without using non-analytic expressions like the square root of a derivative. Since we're concerned with the spectrum of linearized perturbations, the order of derivatives $(u \cdot \partial)$ and $P^\mu_\nu \partial_\mu$ is irrelevant.
The frame transformation incorporates terms of all orders in the derivative expansion, thus, the final transformed stress-energy tensor also have terms of all orders in the derivative expansion.

\smallskip
\textit{Conclusion:--}
In this letter, we have shown that the spectrum of hydrodynamic modes is unaffected by any redefinition of fluid fields, while the spectrum of non-hydrodynamic modes can change under such redefinitions. 
Notably, it is possible to eliminate a non-hydrodynamic mode from the spectrum through an appropriate all-order frame redefinition, as illustrated by our example on the first-order BDNK theory. However, this redefinition cannot fully erase the information about any physical non-hydrodynamic mode. The infinite series of the final stress-energy tensor in the `hydro' frame has a radius of convergence located at the lowest non-hydrodynamic mode in the complex frequency and momentum space\cite{Heller:2013fn,Heller:2020uuy}, suggesting that this mode acts as a cutoff in the hydrodynamic expansion\cite{Grozdanov:2019kge,Heller:2020hnq,Cartwright:2021qpp}.

Our analysis implies that in a causal physical theory, removing physical non-hydrodynamic modes requires including all higher-order terms in the energy-momentum tensor without truncation. In Fourier space, the data concerning the physical non-hydrodynamic mode is encapsulated in the radius of convergence of this infinite series\cite{Bemfica:2019knx,Heller:2022ejw,Gavassino:2021owo,Gavassino:2023myj,Gavassino:2024pgl,Wang:2023csj}. If the energy-momentum tensor is truncated to a finite order, physical non-hydrodynamic modes are necessary for causality. Recent observations indicate that a theory without non-hydrodynamic modes when truncated, may lack causality but remain stable\cite{Hiscock:1985zz,Novak:2019wqg,Armas:2020mpr,Basar:2024qxd,Pu:2009fj}. Our analysis focused on uncharged fluids, where the conservation of the stress tensor is the only hydrodynamic equation of motion. Future work may explore other fluids with various charge conservation equations and MIS-type theories that involve auxiliary fields, such as the shear tensor, which can also be redefined.

We have found that a typical linearized frame redefinition introduces new non-hydrodynamic modes corresponding to the zero eigenspace of the linearized frame transformation. Perturbations with frequencies matching these new modes will vanish after the transformation, trivially satisfying the linearized equations of motion. In the transformed frame, these appear as new modes that modify the UV behavior of the theory without affecting its low-energy physics. We have utilized this characteristic of frame redefinition to construct an equivalent hydrodynamic theory with the non-hydrodynamic modes eliminated from the spectrum of linearized perturbations.
%

\smallskip
\textit{Acknowledgments:--}
S.B., S.M. and S.R. acknowledge the Department of Atomic Energy, India, for the funding support.
R.S. acknowledges kind hospitality and support of INFN Firenze, Department of Physics and Astronomy, University of Florence, ECT* Trento, NCBJ Warsaw, TU Darmstadt, ITP Goethe University where part of this work is completed and is supported partly by postdoctoral fellowship of West University of Timișoara. 
We acknowledge enlightening discussions with L.~Gavassino, M.~Spalinski, and A.~Yarom.
%
\bibliography{pv_ref}{}

\providecommand{\href}[2]{#2}\begingroup\raggedright\begin{thebibliography}{10}

\bibitem{Hiscock:1985zz}
W.~A. Hiscock and L.~Lindblom, ``{Generic instabilities in first-order dissipative relativistic fluid theories},'' \href{http://dx.doi.org/10.1103/PhysRevD.31.725}{{\em Phys. Rev. D} {\bfseries 31} (1985) 725--733}.

\bibitem{Geroch:1990bw}
R.~P. Geroch and L.~Lindblom, ``{Dissipative relativistic fluid theories of divergence type},'' \href{http://dx.doi.org/10.1103/PhysRevD.41.1855}{{\em Phys. Rev. D} {\bfseries 41} (1990) 1855}.

\bibitem{Geroch:1995bx}
R.~P. Geroch, ``{Relativistic theories of dissipative fluids},'' \href{http://dx.doi.org/10.1063/1.530958}{{\em J. Math. Phys.} {\bfseries 36} (1995) 4226}.

\bibitem{Gavassino:2021kjm}
L.~Gavassino, M.~Antonelli, and B.~Haskell, ``{Thermodynamic Stability Implies Causality},'' \href{http://dx.doi.org/10.1103/PhysRevLett.128.010606}{{\em Phys. Rev. Lett.} {\bfseries 128} no.~1, (2022) 010606}, \href{http://arxiv.org/abs/2105.14621}{{\ttfamily arXiv:2105.14621 [gr-qc]}}.

\bibitem{Arnold:2014jva}
P.~Arnold, P.~Romatschke, and W.~van~der Schee, ``{Absence of a local rest frame in far from equilibrium quantum matter},'' \href{http://dx.doi.org/10.1007/JHEP10(2014)110}{{\em JHEP} {\bfseries 10} (2014) 110}, \href{http://arxiv.org/abs/1408.2518}{{\ttfamily arXiv:1408.2518 [hep-th]}}.

\bibitem{Lehner:2017yes}
L.~Lehner, O.~A. Reula, and M.~E. Rubio, ``{Hyperbolic theory of relativistic conformal dissipative fluids},'' \href{http://dx.doi.org/10.1103/PhysRevD.97.024013}{{\em Phys. Rev. D} {\bfseries 97} no.~2, (2018) 024013}, \href{http://arxiv.org/abs/1710.08033}{{\ttfamily arXiv:1710.08033 [gr-qc]}}.

\bibitem{Israel:1976tn}
W.~Israel, ``{Nonstationary irreversible thermodynamics: A Causal relativistic theory},'' \href{http://dx.doi.org/10.1016/0003-4916(76)90064-6}{{\em Annals Phys.} {\bfseries 100} (1976) 310--331}.

\bibitem{Israel:1979wp}
W.~Israel and J.~M. Stewart, ``{Transient relativistic thermodynamics and kinetic theory},'' \href{http://dx.doi.org/10.1016/0003-4916(79)90130-1}{{\em Annals Phys.} {\bfseries 118} (1979) 341--372}.

\bibitem{Bemfica:2017wps}
F.~S. Bemfica, M.~M. Disconzi, and J.~Noronha, ``{Causality and existence of solutions of relativistic viscous fluid dynamics with gravity},'' \href{http://dx.doi.org/10.1103/PhysRevD.98.104064}{{\em Phys. Rev. D} {\bfseries 98} no.~10, (2018) 104064}, \href{http://arxiv.org/abs/1708.06255}{{\ttfamily arXiv:1708.06255 [gr-qc]}}.

\bibitem{Kovtun:2019hdm}
P.~Kovtun, ``{First-order relativistic hydrodynamics is stable},'' \href{http://dx.doi.org/10.1007/JHEP10(2019)034}{{\em JHEP} {\bfseries 10} (2019) 034}, \href{http://arxiv.org/abs/1907.08191}{{\ttfamily arXiv:1907.08191 [hep-th]}}.

\bibitem{Bemfica:2019cop}
F.~S. Bemfica, M.~M. Disconzi, and J.~Noronha, ``{Causality of the Einstein-Israel-Stewart Theory with Bulk Viscosity},'' \href{http://dx.doi.org/10.1103/PhysRevLett.122.221602}{{\em Phys. Rev. Lett.} {\bfseries 122} no.~22, (2019) 221602}, \href{http://arxiv.org/abs/1901.06701}{{\ttfamily arXiv:1901.06701 [gr-qc]}}.

\bibitem{Denicol:2008ha}
G.~S. Denicol, T.~Kodama, T.~Koide, and P.~Mota, ``{Stability and Causality in relativistic dissipative hydrodynamics},'' \href{http://dx.doi.org/10.1088/0954-3899/35/11/115102}{{\em J. Phys. G} {\bfseries 35} (2008) 115102}, \href{http://arxiv.org/abs/0807.3120}{{\ttfamily arXiv:0807.3120 [hep-ph]}}.

\bibitem{Floerchinger:2017cii}
S.~Floerchinger and E.~Grossi, ``{Causality of fluid dynamics for high-energy nuclear collisions},'' \href{http://dx.doi.org/10.1007/JHEP08(2018)186}{{\em JHEP} {\bfseries 08} (2018) 186}, \href{http://arxiv.org/abs/1711.06687}{{\ttfamily arXiv:1711.06687 [nucl-th]}}.

\bibitem{Biswas:2020rps}
R.~Biswas, A.~Dash, N.~Haque, S.~Pu, and V.~Roy, ``{Causality and stability in relativistic viscous non-resistive magneto-fluid dynamics},'' \href{http://dx.doi.org/10.1007/JHEP10(2020)171}{{\em JHEP} {\bfseries 10} (2020) 171}, \href{http://arxiv.org/abs/2007.05431}{{\ttfamily arXiv:2007.05431 [nucl-th]}}.

\bibitem{Hoult:2023clg}
R.~E. Hoult and P.~Kovtun, ``{Causality and classical dispersion relations},'' \href{http://dx.doi.org/10.1103/PhysRevD.109.046018}{{\em Phys. Rev. D} {\bfseries 109} no.~4, (2024) 046018}, \href{http://arxiv.org/abs/2309.11703}{{\ttfamily arXiv:2309.11703 [hep-th]}}.

\bibitem{Sammet:2023bfo}
J.~Sammet, M.~Mayer, and D.~H. Rischke, ``{Linear stability analysis of Israel-Stewart theory in the case of a nonzero background charge},'' \href{http://dx.doi.org/10.1103/PhysRevD.107.114028}{{\em Phys. Rev. D} {\bfseries 107} no.~11, (2023) 114028}, \href{http://arxiv.org/abs/2302.01070}{{\ttfamily arXiv:2302.01070 [hep-th]}}.

\bibitem{Domingues:2024pom}
T.~S. Domingues, R.~Krupczak, J.~Noronha, T.~N. da~Silva, J.-F. Paquet, and M.~Luzum, ``{The effect of causality constraints on Bayesian analyses of heavy-ion collisions},'' \href{http://arxiv.org/abs/2409.17127}{{\ttfamily arXiv:2409.17127 [nucl-th]}}.

\bibitem{Noronha:2021syv}
J.~Noronha, M.~Spali\'nski, and E.~Speranza, ``{Transient Relativistic Fluid Dynamics in a General Hydrodynamic Frame},'' \href{http://dx.doi.org/10.1103/PhysRevLett.128.252302}{{\em Phys. Rev. Lett.} {\bfseries 128} no.~25, (2022) 252302}, \href{http://arxiv.org/abs/2105.01034}{{\ttfamily arXiv:2105.01034 [nucl-th]}}.

\bibitem{Mitra:2023ipl}
S.~Mitra, ``{Acausality in truncated M\"uller-Israel-Stewart-type theory},'' \href{http://dx.doi.org/10.1103/PhysRevD.109.L121501}{{\em Phys. Rev. D} {\bfseries 109} no.~12, (2024) L121501}, \href{http://arxiv.org/abs/2312.10972}{{\ttfamily arXiv:2312.10972 [gr-qc]}}.

\bibitem{Bhattacharyya:2023srn}
S.~Bhattacharyya, S.~Mitra, and S.~Roy, ``{Frame transformation and stable-causal hydrodynamic theory},'' \href{http://arxiv.org/abs/2312.16407}{{\ttfamily arXiv:2312.16407 [nucl-th]}}.

\bibitem{Bhattacharyya:2024tfj}
S.~Bhattacharyya, S.~Mitra, and S.~Roy, ``{Causality and stability in relativistic hydrodynamic theory - a choice to be endured},'' \href{http://dx.doi.org/10.1016/j.physletb.2024.138918}{{\em Phys. Lett. B} {\bfseries 856} (2024) 138918}, \href{http://arxiv.org/abs/2407.18997}{{\ttfamily arXiv:2407.18997 [nucl-th]}}.

\bibitem{Mitra:2024yei}
S.~Mitra, ``{Why do newer degrees of freedom appear in higher-order truncated hydrodynamic theory?},'' \href{http://arxiv.org/abs/2408.15088}{{\ttfamily arXiv:2408.15088 [nucl-th]}}.

\bibitem{Xie:2023gbo}
X.-Q. Xie, D.-L. Wang, C.~Yang, and S.~Pu, ``{Causality and stability analysis for the minimal causal spin hydrodynamics},'' \href{http://dx.doi.org/10.1103/PhysRevD.108.094031}{{\em Phys. Rev. D} {\bfseries 108} no.~9, (2023) 094031}, \href{http://arxiv.org/abs/2306.13880}{{\ttfamily arXiv:2306.13880 [hep-ph]}}.

\bibitem{Grozdanov:2018fic}
S.~Grozdanov, A.~Lucas, and N.~Poovuttikul, ``{Holography and hydrodynamics with weakly broken symmetries},'' \href{http://dx.doi.org/10.1103/PhysRevD.99.086012}{{\em Phys. Rev. D} {\bfseries 99} no.~8, (2019) 086012}, \href{http://arxiv.org/abs/1810.10016}{{\ttfamily arXiv:1810.10016 [hep-th]}}.

\bibitem{Grozdanov:2019uhi}
S.~Grozdanov, P.~K. Kovtun, A.~O. Starinets, and P.~Tadi\'c, ``{The complex life of hydrodynamic modes},'' \href{http://dx.doi.org/10.1007/JHEP11(2019)097}{{\em JHEP} {\bfseries 11} (2019) 097}, \href{http://arxiv.org/abs/1904.12862}{{\ttfamily arXiv:1904.12862 [hep-th]}}.

\bibitem{Bea:2023rru}
Y.~Bea and P.~Figueras, ``{Field redefinitions and evolutions in relativistic Navier-Stokes},'' \href{http://arxiv.org/abs/2312.16671}{{\ttfamily arXiv:2312.16671 [hep-th]}}.

\bibitem{Hoult:2024cyx}
R.~E. Hoult and A.~Shukla, ``{Causal and Stable Superfluid Hydrodynamics},'' \href{http://arxiv.org/abs/2410.22855}{{\ttfamily arXiv:2410.22855 [hep-th]}}.

\bibitem{Heller:2022ejw}
M.~P. Heller, A.~Serantes, M.~Spali\'nski, and B.~Withers, ``{Rigorous Bounds on Transport from Causality},'' \href{http://dx.doi.org/10.1103/PhysRevLett.130.261601}{{\em Phys. Rev. Lett.} {\bfseries 130} no.~26, (2023) 261601}, \href{http://arxiv.org/abs/2212.07434}{{\ttfamily arXiv:2212.07434 [hep-th]}}.

\bibitem{Bhattacharyya:2024ohn}
S.~Bhattacharyya, S.~Mitra, S.~Roy, and R.~Singh, ``{Field redefinition and its impact in relativistic hydrodynamics},'' \href{http://arxiv.org/abs/2409.15387}{{\ttfamily arXiv:2409.15387 [nucl-th]}}.

\bibitem{Heller:2013fn}
M.~P. Heller, R.~A. Janik, and P.~Witaszczyk, ``{Hydrodynamic Gradient Expansion in Gauge Theory Plasmas},'' \href{http://dx.doi.org/10.1103/PhysRevLett.110.211602}{{\em Phys. Rev. Lett.} {\bfseries 110} no.~21, (2013) 211602}, \href{http://arxiv.org/abs/1302.0697}{{\ttfamily arXiv:1302.0697 [hep-th]}}.

\bibitem{Heller:2020uuy}
M.~P. Heller, A.~Serantes, M.~Spali\'nski, V.~Svensson, and B.~Withers, ``{Hydrodynamic gradient expansion in linear response theory},'' \href{http://dx.doi.org/10.1103/PhysRevD.104.066002}{{\em Phys. Rev. D} {\bfseries 104} no.~6, (2021) 066002}, \href{http://arxiv.org/abs/2007.05524}{{\ttfamily arXiv:2007.05524 [hep-th]}}.

\bibitem{Grozdanov:2019kge}
S.~Grozdanov, P.~K. Kovtun, A.~O. Starinets, and P.~Tadi\'c, ``{Convergence of the Gradient Expansion in Hydrodynamics},'' \href{http://dx.doi.org/10.1103/PhysRevLett.122.251601}{{\em Phys. Rev. Lett.} {\bfseries 122} no.~25, (2019) 251601}, \href{http://arxiv.org/abs/1904.01018}{{\ttfamily arXiv:1904.01018 [hep-th]}}.

\bibitem{Heller:2020hnq}
M.~P. Heller, A.~Serantes, M.~Spali\'nski, V.~Svensson, and B.~Withers, ``{Convergence of hydrodynamic modes: insights from kinetic theory and holography},'' \href{http://dx.doi.org/10.21468/SciPostPhys.10.6.123}{{\em SciPost Phys.} {\bfseries 10} no.~6, (2021) 123}, \href{http://arxiv.org/abs/2012.15393}{{\ttfamily arXiv:2012.15393 [hep-th]}}.

\bibitem{Cartwright:2021qpp}
C.~Cartwright, M.~G. Amano, M.~Kaminski, J.~Noronha, and E.~Speranza, ``{Convergence of hydrodynamics in a rotating strongly coupled plasma},'' \href{http://dx.doi.org/10.1103/PhysRevD.108.046014}{{\em Phys. Rev. D} {\bfseries 108} no.~4, (2023) 046014}, \href{http://arxiv.org/abs/2112.10781}{{\ttfamily arXiv:2112.10781 [hep-th]}}.

\bibitem{Bemfica:2019knx}
F.~S. Bemfica, F.~S. Bemfica, M.~M. Disconzi, M.~M. Disconzi, J.~Noronha, and J.~Noronha, ``{Nonlinear Causality of General First-Order Relativistic Viscous Hydrodynamics},'' \href{http://dx.doi.org/10.1103/PhysRevD.100.104020}{{\em Phys. Rev. D} {\bfseries 100} no.~10, (2019) 104020}, \href{http://arxiv.org/abs/1907.12695}{{\ttfamily arXiv:1907.12695 [gr-qc]}}. [Erratum: Phys.Rev.D 105, 069902 (2022)].

\bibitem{Gavassino:2021owo}
L.~Gavassino, ``{Can We Make Sense of Dissipation without Causality?},'' \href{http://dx.doi.org/10.1103/PhysRevX.12.041001}{{\em Phys. Rev. X} {\bfseries 12} no.~4, (2022) 041001}, \href{http://arxiv.org/abs/2111.05254}{{\ttfamily arXiv:2111.05254 [gr-qc]}}.

\bibitem{Gavassino:2023myj}
L.~Gavassino, ``{Bounds on transport from hydrodynamic stability},'' \href{http://dx.doi.org/10.1016/j.physletb.2023.137854}{{\em Phys. Lett. B} {\bfseries 840} (2023) 137854}, \href{http://arxiv.org/abs/2301.06651}{{\ttfamily arXiv:2301.06651 [hep-th]}}.

\bibitem{Gavassino:2024pgl}
L.~Gavassino, ``{Infinite Order Hydrodynamics: An Analytical Example},'' \href{http://dx.doi.org/10.1103/PhysRevLett.133.032302}{{\em Phys. Rev. Lett.} {\bfseries 133} no.~3, (2024) 032302}, \href{http://arxiv.org/abs/2402.19343}{{\ttfamily arXiv:2402.19343 [nucl-th]}}.

\bibitem{Wang:2023csj}
D.-L. Wang and S.~Pu, ``{Stability and causality criteria in linear mode analysis: Stability means causality},'' \href{http://dx.doi.org/10.1103/PhysRevD.109.L031504}{{\em Phys. Rev. D} {\bfseries 109} no.~3, (2024) L031504}, \href{http://arxiv.org/abs/2309.11708}{{\ttfamily arXiv:2309.11708 [hep-th]}}.

\bibitem{Novak:2019wqg}
I.~Novak, J.~Sonner, and B.~Withers, ``{Hydrodynamics without boosts},'' \href{http://dx.doi.org/10.1007/JHEP07(2020)165}{{\em JHEP} {\bfseries 07} (2020) 165}, \href{http://arxiv.org/abs/1911.02578}{{\ttfamily arXiv:1911.02578 [hep-th]}}.

\bibitem{Armas:2020mpr}
J.~Armas and A.~Jain, ``{Effective field theory for hydrodynamics without boosts},'' \href{http://dx.doi.org/10.21468/SciPostPhys.11.3.054}{{\em SciPost Phys.} {\bfseries 11} no.~3, (2021) 054}, \href{http://arxiv.org/abs/2010.15782}{{\ttfamily arXiv:2010.15782 [hep-th]}}.

\bibitem{Basar:2024qxd}
G.~Ba\c{s}ar, J.~Bhambure, R.~Singh, and D.~Teaney, ``{Stochastic relativistic advection diffusion equation~from the Metropolis algorithm},'' \href{http://dx.doi.org/10.1103/PhysRevC.110.044903}{{\em Phys. Rev. C} {\bfseries 110} no.~4, (2024) 044903}, \href{http://arxiv.org/abs/2403.04185}{{\ttfamily arXiv:2403.04185 [nucl-th]}}.

\bibitem{Pu:2009fj}
S.~Pu, T.~Koide, and D.~H. Rischke, ``{Does stability of relativistic dissipative fluid dynamics imply causality?},'' \href{http://dx.doi.org/10.1103/PhysRevD.81.114039}{{\em Phys. Rev. D} {\bfseries 81} (2010) 114039}, \href{http://arxiv.org/abs/0907.3906}{{\ttfamily arXiv:0907.3906 [hep-ph]}}.

\end{thebibliography}\endgroup
\bibliographystyle{utphys}
\end{document}